\documentclass[%
 reprint,
superscriptaddress,
 amsmath,amssymb,
 aps,
 prl
]{revtex4-1}

\usepackage{graphicx}
\usepackage{dcolumn}
\usepackage{bm}
\usepackage{upgreek}
\usepackage[colorlinks=true, citecolor=blue,allcolors=blue]{hyperref}

\newcommand{\Oxygen}{{\textsc{O}}}
\newcommand{\Carbon}{{\textsc{C}}}
\newcommand{\Atom}{{\textsc{A}}}

\begin{document}

\preprint{APS/123-QED}

\title{Attosecond Intramolecular-Scattering and Vibronic Delays}

\author{Bejan Ghomashi}
\affiliation{University of Colorado at Boulder, Boulder, Colorado 80309, USA}

\author{Nicolas Douguet}%
\affiliation{Department of Physics, Kennesaw State University, Marietta, GA, USA}

\author{Luca Argenti}
\email{luca.argenti@ucf.edu}
\affiliation{Department of Physics \& CREOL, University of Central Florida, Orlando, Florida 32816, USA}

\date{\today}

\begin{abstract}
The photoionization of the CO molecule from the C$-1s$ orbital does not obey the Franck-Condon approximation, as a consequence of the nuclear recoil that accompanies the direct emission and intra-molecular scattering of the photoelectron. We use an analytical model to investigate the temporal signature of the entangled nuclear and electronic motion in this process. We show that the photoelectron emission delay can be decomposed into its localization and resonant-confinement components. Finally, photoionization by a broadband soft-x-ray pulse results in a coherent vibrational ionic state with a tunable delay with respect to the classical sudden-photoemission limit.
\end{abstract}

\maketitle
The advances in the generation of ultrafast and tunable light sources \cite{Sansone443,Ferrari2010,Chini2014,PhysRevLett.109.083001,CHEN2010173901} have allowed us to explore quantum phenomena beyond the limits of femtochemistry \cite{Zewail2000}, by accessing electron dynamics in atoms and molecules at its natural time scale. While numerous studies have already considered various aspects of the photoionization time delay in atoms \cite{Schultze2010a,Isinger893,Cirelli18}, the focus of photoionization chronoscopy has recently moved to polyatomic systems, from simple molecules~\cite{Vos18,Chacon14,Serov16,Baykusheva17,Huppert16,Hockett16} to complex organic compounds~\cite{Beaulieu18,Beaulieu17}. These studies contribute to outline the perimeter of the new discipline of attochemistry~\cite{Nisoli19}, in which attosecond spectroscopic techniques are used for real-time control of chemical reactions \cite{Nisoli17}, with the long-term aim of applying them to systems of biological, technological, and medical relevance. 

Photoionization by an attosecond pulse can be used to trigger charge migration in a biomolecule, a process that can be tracked in real time, e.g., by using a delayed IR probe pulse to induce the dissociation of the molecule, combined with the detection of cationic fragments \cite{Calegari336}. The lifetime of charge migration is dictated not only by the residual coherence of multiple electronic states of the ion \cite{PhysRevLett.106.053003,Lara-Astiaso16}, but also by the vibrational degrees of freedom \cite{PhysRevLett.121.203002,Despre15}. Understanding the interplay between the electron dynamics initiated by sudden photoionization and the slower nuclear motion, therefore, is essential to describe the evolution of a molecule in the few femtoseconds that follow its ionization by an attosecond pulse.
\begin{figure}[hbtp!]
	\includegraphics[width=\columnwidth]{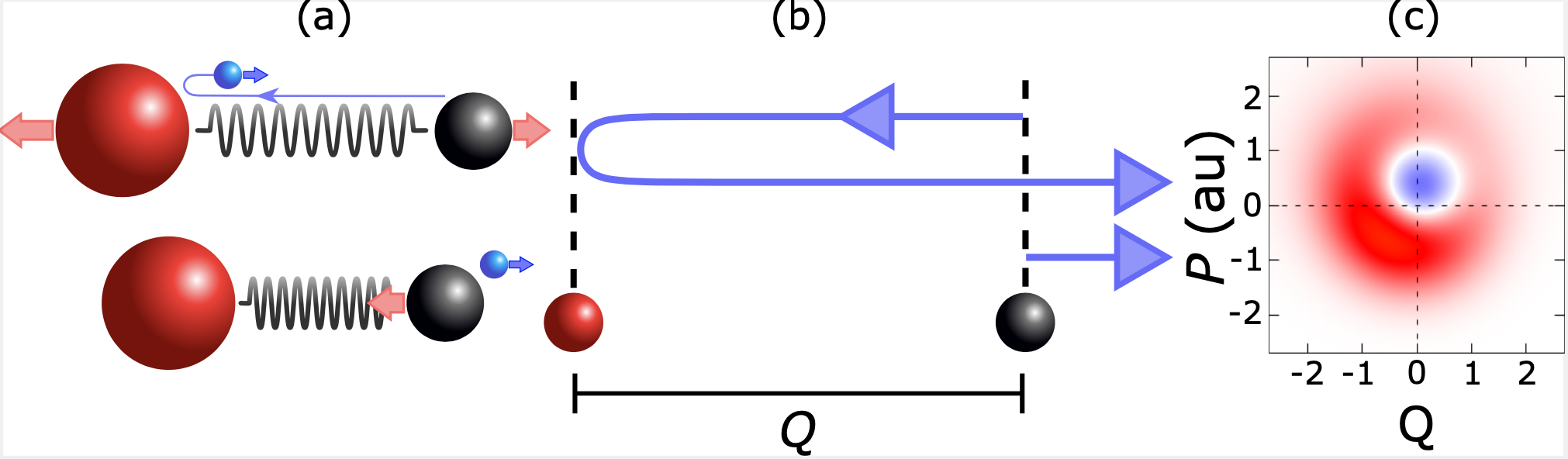}
	\caption{\label{fig:1}(a) Pictorial representation of the nuclear recoil as the photoelectron from the C-$1s$ ionization of CO is emitted towards the other nucleus, possibly reflecting many times ({top}), or directly out ({bottom}), resulting in an expansive and compressive boost, respectively. (b) Representation of two interfering pathways of the outgoing photoelectron. (c) The corresponding ionic nuclear Wigner Distribution when the molecule is ionized by an ultrashort pulse, showing that the ion emerges in a non-classical coherent state.}
\end{figure}

When a molecule is ionized by an x-ray photon from a localized core orbital, the emerging photoelectron collides with nearby nuclei giving rise to the well-known interference patterns of EXAFS spectroscopy~\cite{Lytle1999}. Vibrationally-resolved spectra bear the signature of the energy transferred to the nuclei by the intramolecular scattering process~\cite{Plesiat2012,Ueda2013,Kukk2013,Patanen2014}. New pulsed x-ray sources, such as XFELS~\cite{Pellegrini2016,Seddon2017}, make it now possible to study this phenomenon resolved in time. In this work, we simulate the real-time dynamics of the CO molecule following the C-1s ionization induced by a coherent soft-x-ray pulse using a simplified 1D theoretical model. The vibrationally-resolved photoemission delay results from the interplay between two different phenomena: the localization of the photoelectron at its birth and its resonant confinement by the two nuclei. A short pulse creates ions in partially coherent vibrational states, either in compression or in expansion, depending on the pulse central energy relative to confinement resonances. The vibrational delay, due to intramolecular scattering, is imprinted in the deviation of the nuclear Wigner distribution (WD) from the sudden-photoemission limit. 

The vibrational excitations that accompany photoionization at low electron energy usually follow the Franck-Condon (FC) principle~\cite{Franck1926,Condon1928}, which states that  the positions and momenta of the nuclei remain unchanged during ionization. 
In x-ray photoionization, however, the photoelectron may be ejected from a localized core orbital with an energy of several hundred electronvolts, thus resulting in a large nuclear recoil 
that causes the expansion or contraction of the molecule, as exemplified in \hbox{Fig. \ref{fig:1}} for the C-$1s$ photoionization of carbon monoxide. On top of direct photoemission, the molecule is also excited by the nuclear recoil associated to the photoelectron scattering off neighboring nuclei~\cite{Kukk2013}. These multiple ionization pathways interfere, leading to noticeable effects in both the photoelectron amplitude and the vibrational state of the ion, represented by its vibrational WD. Multiple recoils give rise to a transient confinement of the photoelectron by the molecular structure, which manifests itself as a series of broad high-energy shape resonances, equally spaced in momentum. 


Let's approximate a CO molecule excited by an attosecond pulse with a 1D model Hamiltonian in the velocity gauge and in the dipole approximation,
\begin{equation}
H=\frac{P_R^2}{2\mu}+\frac{\mu\Omega^2}{2}(R-R_0)^2 + \frac{p^2}{2m}+V_e(x;R) +  A(t) p,
\end{equation}
where $R$ is the internuclear distance, $P_R$ is its conjugate momentum, with $\mu=m_\Carbon m_\Oxygen/(m_\Carbon+m_\Oxygen)$ is the reduced mass of the system, $x$ and $p$ are the position and momentum of the photoelectron, respectively, and $A(t)$ is the vector potential of the external field, polarized along the molecular axis. Atomic units ($m_e=1$, $\hbar=1$, $e=1$) are used throughout, unless otherwise stated. The vibrational motion of the molecule is described by a harmonic oscillator with frequency $\Omega=2130\ \rm{cm}^{-1}$ 
and equilibrium position $R_0=2.13$ a.u. (1.13 \AA)~\cite{Kukk2013}. The 1D electronic potential $V_e(x;R)$, which depends parametrically on $R$, is assumed to have the elementary form
\begin{equation}
V_e(x;R)=V_\Oxygen+V_\Carbon,\qquad V_\Atom = v_\Atom\,\delta(x-R_\Atom)
\end{equation}
where $R_\Carbon=\mu R/m_\Carbon$ and $R_\Oxygen=-\mu R/m_\Oxygen$ are the equilibrium positions of the carbon and oxygen nuclei relative to the center of mass. The constants $v_\Oxygen=-6.3$~a.u. and $v_\Carbon=-4.6$~a.u. are chosen to reproduce the oxygen and carbon core orbital energies of $-19.5$~a.u. and $-10.5$~a.u., respectively. While we could have used a more sophisticated potential to describe the electron dynamics, we chose this simple form since it reproduces the main effects of the nuclear recoil, of the intra-molecular photoelectron scattering, and of the photoelectron resonant confinement. Furthermore, it allows us to decompose the contributions to the vibronic photoemission delay in analytical form.
Within the Born-Oppenheimer approximation, the electron scattering states $|E_\beta;R\rangle$ in the model potential $V_e$, satisfying incoming boundary conditions, can be obtained in closed form by solving the Lippman-Schwinger equation~\cite{Newton},
\begin{align}
|E_\beta;R\rangle=|E_\beta\rangle_0+G_0^-(E)V_e(x;R)|E_\beta;R\rangle,
\end{align}
where $|E_\beta\rangle_0$ is a free scattering state, $G_0^-(E)=(E-p^2/2-i0^+)^{-1}$ is the anticipated resolvent, $E$ is the asymptotic photoelectron energy, and $\beta$ specifies the left or right outgoing character of the scattering states.
The solution can also be expressed as a Born series (we omit the $x$ and $R$ dependence for simplicity),
\begin{equation}\label{eq:Born}
|E_\beta\rangle=\sum_n\left[G_0^-(E)\,(V_\Oxygen+V_\Carbon)\right]^n|E_\beta\rangle_0.
\end{equation}
From~\eqref{eq:Born}, it is possible to differentiate the components of the final wave function that involve only the interaction with the carbon atom from the remaining components. The former are responsible for the nuclear recoil in direct photoemission, whereas the latter, which involve scattering by both nuclei, are responsible for the resonant photoelectron confinement. The series in~\eqref{eq:Born} comprises arbitrary finite sequences of multiple interactions with the two atoms, $\mathsf{c}=G_0^-(E)\,V_\Carbon$ and $\mathsf{o}=G_0^-(E)\,V_\Oxygen$, e.g., $\mathsf{ccocooocc}$, etc.
Since the scattering off an isolated singular potential is known, it is convenient to carry out a resummation of the series in~\eqref{eq:Born} to arbitrary order in the interaction with either one or the other of the two nuclei, $\mathsf{C}=\mathcal{G}^-_\Carbon(E) V_\Carbon = \mathsf{c}+\mathsf{cc}+\mathsf{ccc}+\ldots$, where $\mathcal{G}^-_\Carbon(E)=(E-H_0-V_\Carbon-i0^+)^{-1}$, and $\mathsf{O}=\mathcal{G}^-_\Oxygen(E) V_\Oxygen$. The solution can then be expressed in the closed form
$|E_\beta\rangle=(1+\mathsf{C})(1-\mathsf{OC})^{-1}(1+\mathsf{O})|E_\beta\rangle_0$.
To analyze the photoemission from the C-$1s$ orbital to the carbon side ($\beta=C$), we split the expression in a direct-emission background component, where the effect of the oxygen atom is ignored, $|E_\mathrm{C},\mathsf{bg}\rangle = (1+\mathsf{C})|E_\mathrm{C}\rangle_0$, and a residual resonant component, which accounts for the arbitrary number of intra-molecular scattering events the photoelectron can undergo prior to leaving the molecule, $|E_\mathrm{C},\mathsf{res}\rangle =(1+\mathsf{C})\, (1-\mathsf{OC})^{-1}\mathsf{O}|E_\mathrm{C},\mathsf{bg}\rangle$. The direct ionization from the carbon atom and away from the oxygen atom is accompanied by a recoil that compresses the molecule. In the resonant component, the photoelectron recoils inward, thus boosting the nuclei outward.
\begin{figure}[hbtp!]
	\includegraphics[width=\columnwidth]{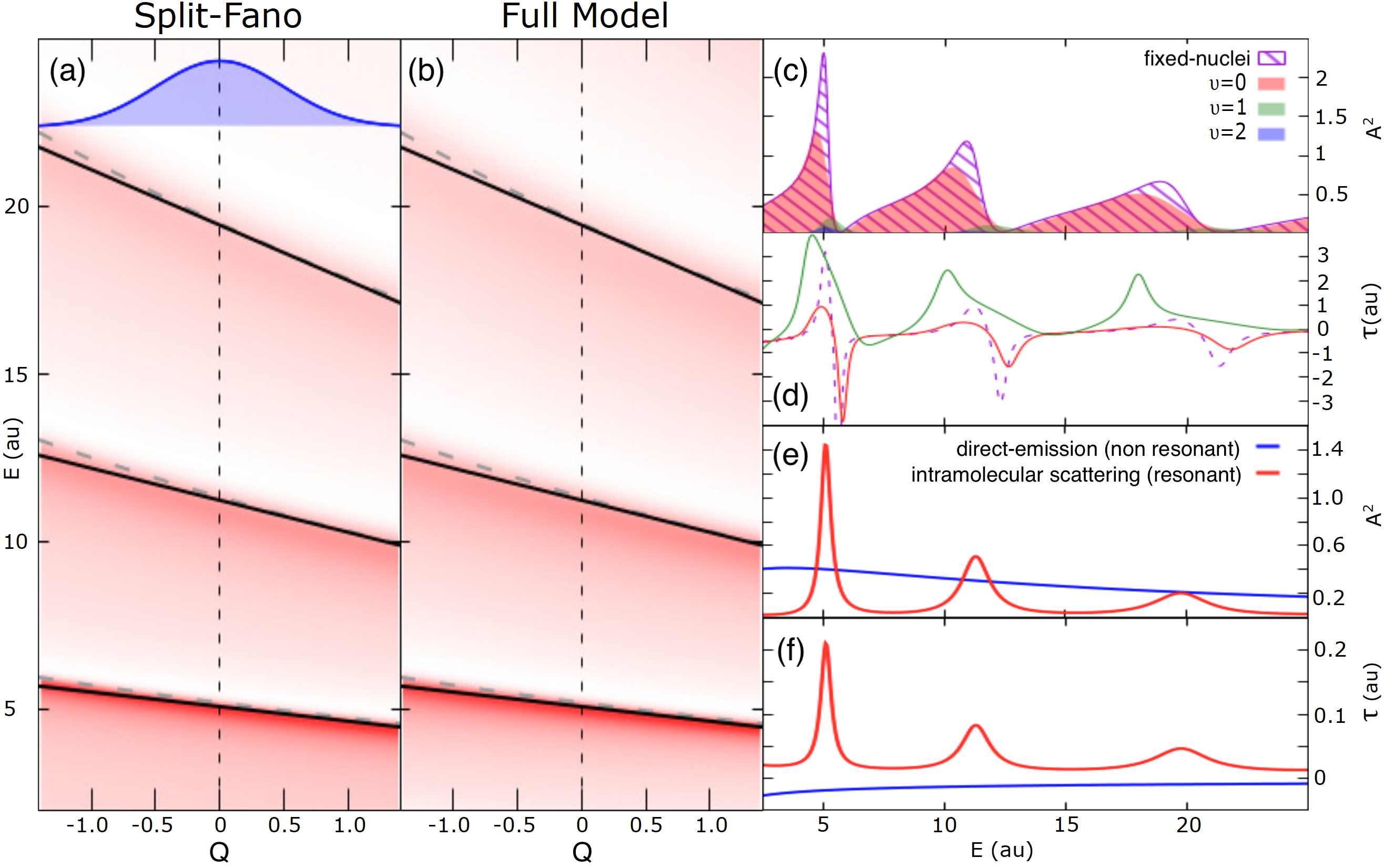}
	\caption{\label{fig:vib-delay} (a-b) Square module of the dipole transition amplitude $|\mu_{C,R}^-(E)|^2$ in the parametrized Fano model (a) and in the exact calculation (b). The particle-in-the-box energies (grey dashed lines) follow closely the confinement resonances, and are approximated well by linear tangents (black solid lines). (c) Fixed-nuclei $|\mu_{C,R_0}^-(E)|^2$ (electronic), and vibrationally resolved $|\mu_{C,\nu}^-(E)|^2$. (d) Fixed-nuclei (dashed line) and vibrationally resolved photoemission delays. (e) Square module and (f) time delays of the expansion (intramolecular scattering) and compression (direct ionization) components of the fixed-nuclei photoionization amplitude.}
\end{figure}

The one-photon ionization amplitude from the C-$1s$ orbital of the CO molecule in its vibrational ground state to a given vibrational state $|\nu\rangle$ of the parent ion is
\begin{equation}
\mathcal{A}^{-}_{\beta,\nu}(E)=\mu^-_{\beta,\nu}(E)\tilde{A}(E+\Omega\nu-E_C),
\label{eq:one-photon-ampl}
\end{equation}
where $E$ is the photoelectron energy and $-E_\Carbon$ is the ionization potential. 
In~\eqref{eq:one-photon-ampl}, $\mu^-_{\beta,\nu}(E)=\langle\nu|\mu^-_{\beta,R}(E)|0\rangle_R$, where $\mu^-_{\beta,R}(E)=\langle E,\beta;R|p|\varphi_C\rangle_x$ is 
the fixed-nuclei transition dipole moment, $|\varphi_C\rangle$ is the C$-1s$ orbital,
and $\tilde{A}(\omega)=\int dt A(t) e^{i\omega t}$ is the Fourier transform of the vector potential. 
Figure~\ref{fig:vib-delay}b shows $|\mu^-_{\beta,R}(E)|^2$ as a function of the C-O bond length and of the photoelectron energy, computed using the full analytical model. The molecule exhibits characteristic Fano profiles associated to the shape resonances of the electron confined by the two atoms, $\mu^-_{\mathsf{\beta},R}(E)\approx \sum_{n}\frac{\epsilon_n(R)+q_n}{\epsilon_n(R)-i}$, where $\epsilon_n(R)=2[E-E_n(R)]/\Gamma_n$ is the reduced photoelectron energy relative to the $n-$th resonance, with energy $E_n(R)$ and width $\Gamma_n$, and $q_n$ is the asymmetry parameter of the resonance. Figure~\ref{fig:vib-delay}a shows the dipole parametrized with the Fano formula. 
Since the dependence of $q_n$ and $\Gamma_n$ on $R$ is not crucial, we neglect it, keeping the value of $q_n$ and $\Gamma_n$ at equilibrium, $q_n=q_n(R_0)$ and $\Gamma_n=\Gamma_n(R_0)$. The virtually perfect agreement between the two panels confirms the generality of Fano's formula for resonant photoionization amplitudes. 

While in the real CO$^+$ C$-1s^{-1}$ ion the equilibrium bond length is shorter and the vibrational frequency higher than in the neutral molecule, 
here we use the same harmonic potential for both the neutral and the ion to highlight the effect of the vibrational excitations due solely to recoil. 
Figure~\ref{fig:vib-delay}c shows the fixed-nuclei $|\mu^-_{\beta,R_o}(E)|^2$ near the second, third, and fourth confinement resonances, and vibrationally resolved, $|\mu^-_{\beta,\nu}(E)|^2$, for $\nu=0$, 1, 2. From the $\nu=1$ amplitude, it is clear that the vibrational excitation has a strong resonant component. In Figure~\ref{fig:vib-delay}d we show the fixed-nuclei Wigner time delay, $\uptau_\textsc{w}(R)=d\arg[\mu^-_{\beta,R}(E)]/dE$, and the vibronic delay, i.e., the group delay for the photoelectron emission associated to the excitation of the ion to a specific vibrational state:
\begin{equation}
\uptau^{\nu\leftarrow 0}_\textsc{w}=\frac{d\arg[\mu^-_{\beta,\nu}(E)]}{dE}.
\end{equation}
The emission delays are highly structured, with both positive and negative peaks in proximity of 
the resonant features, due to the interplay between resonant and non-resonant photoemission amplitude, which gives rise to a destructive interference on either the front or the tail of the photoelectron wave packet~\cite{Chu2010,Argenti2017}. 
We can analyze this effect by separately computing the amplitude and delay associated to the direct photoemission and to the intramolecular-scattering photoemission amplitudes. 
In Figure~\ref{fig:vib-delay}e-f we show the photoionization cross section and time delay, respectively, in the fixed-nuclei case. 
Since the electron originates at a distance $\mu R/m_\Carbon$ from the center of mass, the direct-photoemission component should exhibit a non-resonant negative delay $\tau_{\textsc{dp}}=-\frac{\mu}{m_\Carbon} \frac{R}{\sqrt{2E}}$, which does indeed coincide with the background time-delay curve. 
The amplitude squared, which is quintessentially resonant, has a Lorentzian profile in close proximity to each resonance, as expected. 

The vibronic photoelectron emission delays determine the absolute phase of the total wavefunction of the neutral molecule and hence, for any given outgoing photoelectron wavepacket, the amplitude and phase of the corresponding vibrational components of the ion. When the molecule is ionized by an x-ray pulse with duration shorter than the vibrational period, therefore, the ion emerges in a partially coherent state.
\begin{figure}[hbtp!]
	\includegraphics[width=\columnwidth]{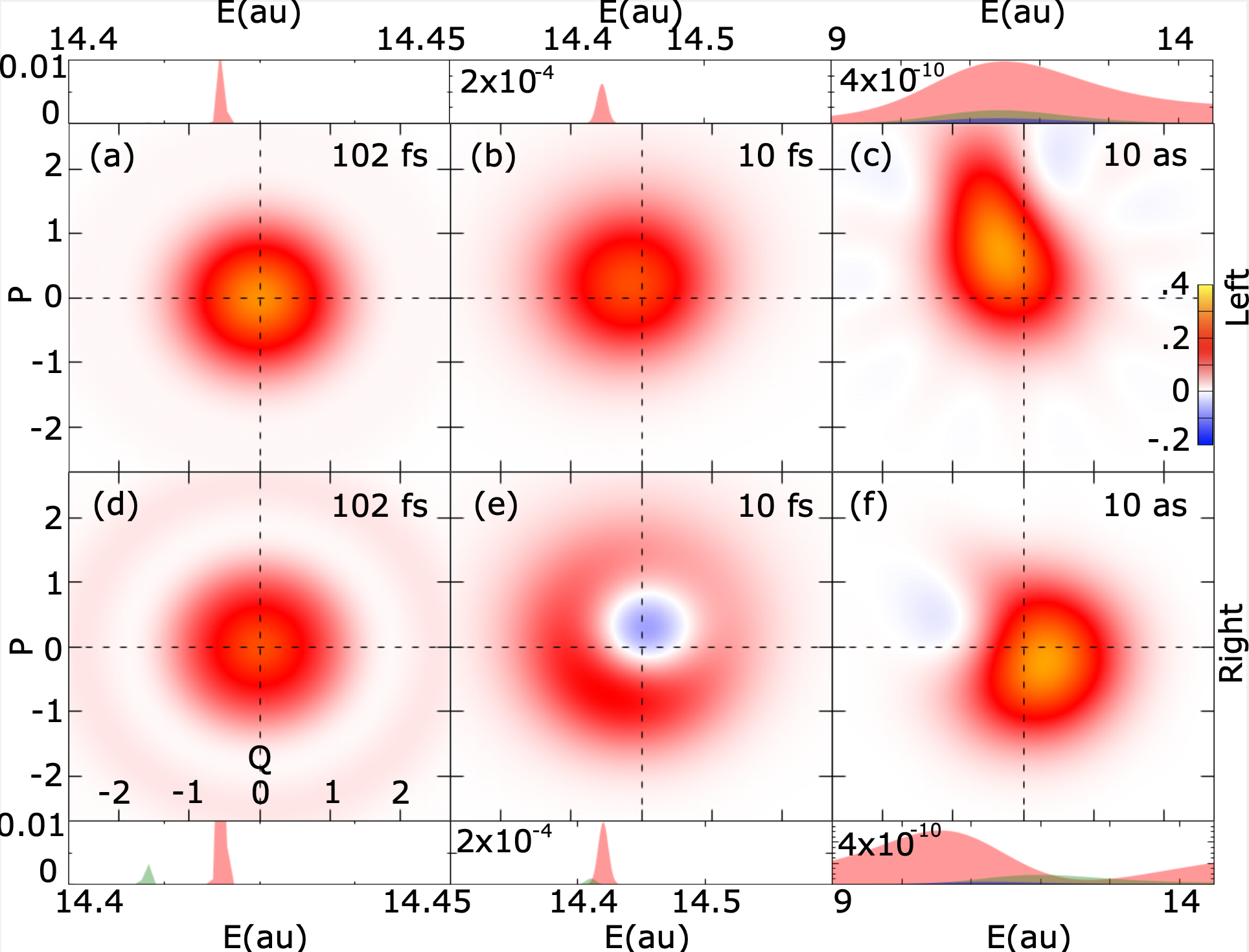}
	\caption{\label{fig:wigner-left}The photoelectron probability ({top} and {bottom} panels) and WD of the parent ion (a-f) for the photoelectron leaving to the left (oxygen side, {top}, a-c) and to the right (carbon side, {bottom}, d-f), from long (left column) to short (right column) pulses. The vibrationally-resolved peaks are shown in solid color ($\nu=0$ red; $\nu=1$ green; $\nu=2$ blue). For long pulses, the ion emerges in an incoherent state. As the pulse shortens, the ion exhibits vibrational coherence, eventually approaching the classical sudden-emission limit.}
\end{figure}
The top panels of Fig.~\ref{fig:wigner-left} show the vibrationally-resolved photoelectron spectra using a soft-x-ray pulse with duration larger, comparable to, and shorter than the vibrational period. In the latter case, all the vibrationally resolved peaks overlap. 
The parent ion, therefore, can retain some coherence even if the energy of the photoelectron is not measured in coincidence. The residual dynamics of the nuclei is visible in the final WD of the vibrational state, in interaction representation
\begin{equation}
    \rho^{(\beta)}(R,P_R)=\sum_{\nu\nu'}\rho^{(\beta)}_{\nu\nu'}\,W_{\nu\nu'}\left[\sqrt{2\mu\Omega}(-R+iP_R/\mu\Omega)\right],
\end{equation}
where 
$\rho^{(\beta)}_{\nu\nu'}=\int dE\, \mathcal{A}^-_{\beta\nu}(E)\mathcal{A}^{-*}_{\beta\nu'}(E)$ is the vibrational density matrix and
\begin{align}
    W_{\nu,\nu'}(\alpha)=2(-1)^{\nu+\nu'}\sqrt{\nu!\nu'!}\,(\alpha^*)^{\nu}\alpha^{\nu'}\,e^{-\frac{1}{2}|\alpha|^2}\times\nonumber\\
\times\sum^{\min(\nu,\nu')}_{\lambda=0}\frac{(-|\alpha|^2)^{-\lambda}}{\lambda!(\nu-\lambda)!(\nu'-\lambda)!}.
\end{align}
Figure~\ref{fig:wigner-left}a-f show the WD as a function of the non-dimensional displacement from equilibrium $Q=(R-R_0)/\sqrt{\omega}$ and the canonically conjugated momentum $P=\sqrt{\omega}P_R$, for emission on either the oxygen (a-c) or carbon (d-f) side. For long pulses, the WD is centered at the origin, as expected, since the ionic state is fully incoherent. As the pulse shortens, however, the distribution barycenter shifts away from the origin, and the distribution assumes non-classical negative values. The vibrational wave packet at the time of ionization is skewed at positive $Q$ and negative $P$ values, i.e., stretched out and contracting. If the photoelectron were to interact only with the ion it originates from and for a time much shorter than the vibrational period, we would expect the vibrational distribution to be centered at $Q=0$, with a positive classical boost $\sqrt{2\omega E}$. Since the WD rotates in time with angular frequency $\omega$, therefore, we can interpret the angular distance $\phi$ of its barycenter from the positive $P$ axis as a vibronic delay $\uptau_\textsc{VIB}=\phi/\omega$.
When the electron leaves on the C side, we can recognize the classical compression boost as the blurry halo in the lower half plane. The dominant feature, however, is an expansion, which is a consequence of intramolecular scattering: the electron, initially moving to the left, bounces off the left-most nucleus and heads out back to the right. 

\begin{figure}
	\includegraphics[width=\columnwidth]{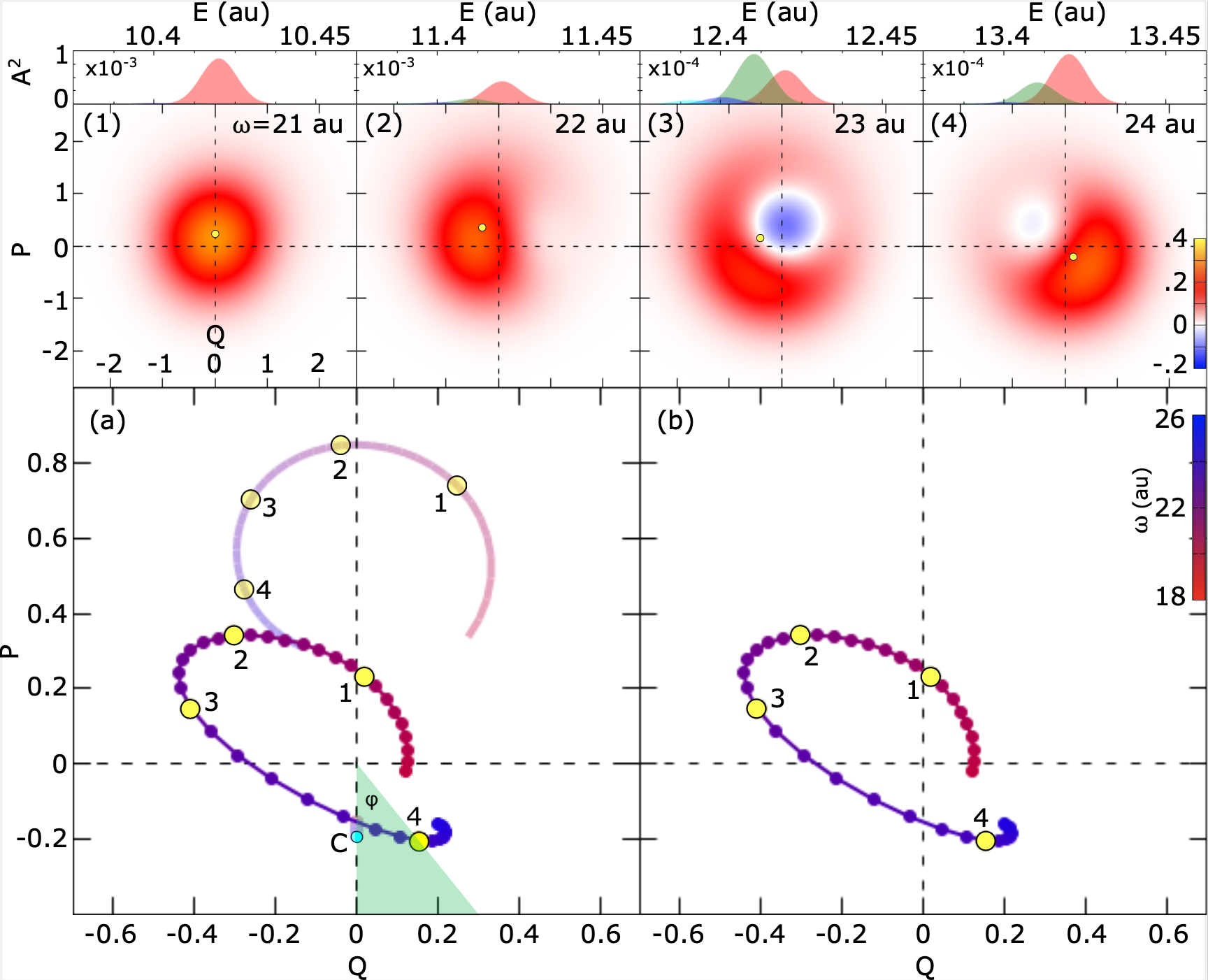}
	\caption{Photoelectron probability ({top}) and WD of the parent ion (1-4) in the case of the photoelectron leaving to the right (carbon side), for central pulse frequencies increasing between two confinement resonances: $\omega_{\textsc{x-ray}}=$21~a.u., 22~a.u., 23~a.u., and 24~a.u. (a) Trajectory of the $(Q,P)$ expectation value as a function of $\omega_{\textsc{x-ray}}$. The light-colored arc is the trajectory of the resonant intramolecular-scattering component. Point C is the background direct-emission component, which coincides with the classical sudden-emission limit. $\phi/\omega$ quantifies the apparent delay of the ionic vibrational state with respect to the classical limit. (b) Same as (a), but using a Fano model for the dipole amplitude, instead of the exact calculation.}
	\label{fig:wigner-right}
\end{figure}

Let us focus now on the WD for the C-side emission as the central frequency of the x-ray wave packet traverses the minimum of the profile next to the fourth confinement resonant, around $\omega_{\textsc{x-ray}}\sim$~22~a.u. The WD has a positive boost at $\omega_{\textsc{x-ray}}=21$~a.u. (see Fig.~\ref{fig:wigner-right}). As the energy increases, its barycenter undergoes a loop around the origin, while its shape gets drastically distorted, acquiring a prominent negative minimum across the I and II quadrants at $\omega_{\textsc{x-ray}}=23$~a.u., before recovering the symmetric shape of a coherent state at $\omega_{\textsc{x-ray}}=25$~a.u. The two lowest panels show the excursion of the barycenter computed either numerically, or analytically, with a simple Fano model. The exquisite agreement between the two calculations, together with the pervasiveness and generality of the Fano model, suggests that these phenomena can be parametrized with simple models in real systems as well. Alongside the trajectory of the barycenter, computed with the full model, we show also the trajectory computed for the direct ionization case (background compression terms) and for the intramolecular scattering case (resonant expansion terms). With only direct ionization, the trajectory is a fixed point that coincides with the recoil boost (point C). The resonant component performs a trajectory that is approximately circular. 
At $\omega_{\textsc{x-ray}}=20$ a.u., the ion excitation is negligible, i.e., the photoelectron emission is recoil-free. This non-classical behavior is due to a recoil cancellation between direct and resonant ionization paths reminiscent of the M{\"o}ssbauer effect~\cite{Mossbauer1958}. 

In conclusion, we have used an exactly solvable 1D molecular ionization model to illustrate new non-adiabatic temporal observables associated to the interplay between electronic and nuclear motion. We have shown that complex vibrationally resolved photoemission delays can be decomposed in a particle-like background localization delay and a wave-like resonant confinement delay, associated to molecular compression and expansion, respectively, as the electron recoils off the nuclei. At selected energies, adjacent confinement resonances can be entangled via the vibrational state resulting in either negative or positive photoemission delay peaks. Moreover, we have shown that ultrashort pulses result in controllable coherent ionic vibrational states that exhibit a delay with respect to the sudden-photoemission approximation (vibronic delay), due to intemolecular photoelectron scattering. These phenomena, which are beyond the reach of traditional single-photon spectroscopies, can be accessed with extensions of attosecond interferometric spectroscopies to the soft-x-ray domain.

L.A.~was supported by the United States National Science Foundation under NSF grant No.~PHY-1607588, by the DOE CAREER grant No.~DE-SC0020311

\bibliography{biblio}

\end{document}